\documentclass[a4paper]{jpconf}
\usepackage{graphicx}
\usepackage{color}
\begin{document}
\title{{\it Ab initio} nuclear structure - the large sparse matrix eigenvalue problem}

\author{James P. Vary$^1$, Pieter Maris$^1$, Esmond Ng$^2$, Chao Yang$^2$, 
Masha Sosonkina$^3$}

\address{
$^1$Department of Physics, Iowa State University, Ames, IA, 50011 USA \\
$^2$Computational Research Division, Lawrence Berkeley National Laboratory, Berkeley, CA  94720, USA\\
$^3$Scalable Computing Laboratory, Ames Laboratory, Iowa State University, Ames, IA, 50011, USA
}

\ead{jvary@iastate.edu}

\begin{abstract}
The structure and reactions of light nuclei represent fundamental and
formidable challenges for microscopic theory based on realistic strong
interaction potentials.  Several {\it ab initio} methods have now
emerged that provide nearly exact solutions for some nuclear
properties. The {\it ab initio} no core shell model (NCSM) and the no
core full configuration (NCFC) method, frame this quantum
many-particle problem as a large sparse matrix eigenvalue problem
where one evaluates the Hamiltonian matrix in a basis space consisting
of many-fermion Slater determinants and then solves for a set of the
lowest eigenvalues and their associated eigenvectors.  The resulting
eigenvectors are employed to evaluate a set of experimental quantities
to test the underlying potential. 
For fundamental problems of interest, the matrix dimension often
exceeds $10^{10}$ and the number of nonzero matrix elements may
saturate available storage on present-day leadership class facilities.
We survey recent results and advances in solving this large sparse
matrix eigenvalue problem. We also outline the challenges that lie
ahead for achieving further breakthroughs in fundamental nuclear
theory using these {\it ab initio} approaches.
\end{abstract}

\section{Introduction}
The structure of the atomic nucleus and its interactions with matter
and radiation have long been the foci of intense theoretical research
aimed at a quantitative understanding based on the underlying strong
inter-nucleon potentials.  Once validated, a successful approach
promises predictive power for key properties of short-lived nuclei
that are present in stellar interiors and in other nuclear
astrophysical settings.  Moreover, new medical diagnostic and
therapeutic applications may emerge as exotic nuclei are predicted and
produced in the laboratory.  Fine tuning nuclear reactor designs to
reduce cost and increase both safety and efficiency are also possible
outcomes with a high precision theory.

Solving for nuclear properties with the best available nucleon-nucleon
(NN) potentials, supplemented by 3-body (NNN) potentials as needed,
using a quantum many-particle framework that respects all the known
symmetries of the potentials is referred to as an "{\it ab initio}"
problem and is recognized to be computationally hard.  Among the few
{\it ab initio} methods available for light nuclei beyond atomic
number $A = 10$, the no core shell model (NCSM) \cite{NCSM12} and the
no core full configuration (NCFC) \cite{Maris09} methods frame the
problem as a large sparse matrix eigenvalue problem one of the focii
of the SciDAC-UNEDF program.

\section{Nuclear physics goals}
Many fundamental properties of nuclei are poorly understood today.  By
gaining insight into these properties we may better utilize these
dense quantum many-body systems to our advantage.  A short
list of outstanding unsolved problems usually includes the following
questions.

\begin{itemize}
\item What controls nuclear saturation - the property that the central
density is nearly the same for all nuclei?
\item How does a successful shell model emerge from the underlying
theory, including predictions for shell and sub-shell closures? 
\item What are the properties of neutron-rich and proton-rich nuclei,
which will be explored at the future FRIB (Facility for Rare Isotope Beams)?
\item How precise can we predict the properties of nuclei - will it be
feasible
to use nuclei as laboratories for tests of fundamental symmetries in
nature?
\end{itemize}

The past few years have seen substantial progress but the complete
answers are yet to be achieved.  We highlight some of the recent
physics achievements of the NCSM that encourage us to believe that
those major goals may be achievable in the next 10 years.  In many
cases, we showed the critical role played by 3-body forces and the
need for large basis spaces to obtain agreement with experiment.

Since 3-body potentials enlarge the computational requirements by one
to two orders of magnitude in both memory and CPU time, 
we have also worked to demonstrate that
similar results may be achieved by suitable adjustments of the
undetermined features of the nucleon-nucleon (NN) potential, known as
the off-shell properties of the potential. The adjustments were
constructed so as to preserve the original high precision description
of the NN scattering data. A partial list of the NCSM and NCFC
achievements is included below.


\begin{itemize}
\item Described the anomaly of the nearly vanishing quadrupole moment
of $^6$Li \cite{Navratil01};
\item Established need for 3-body potentials to explain, among other
properties, neutrino-$^{12}C$ cross sections \cite{3body_a};
\item Found quenching of Gamow-Teller (GT) transitions in light nuclei
due to effects of 3-body potential
\cite{Negret06} (or off-shell modifications of NN potential) plus
configuration mixing \cite{Vaintraub09};
\item Obtained successful description of $A=10$ to $13$ low-lying
spectra with chiral NN + NNN interactions \cite{Navratil07};
\item Explained ground state spin of $^{10}$B by including chiral NNN
interaction \cite{Navratil07}.
\end{itemize}

We have identified several major near-term physics goals
on the path to the larger goals listed above.  These include:
\begin{itemize}
\item Evaluate the properties of the Hoyle state in $^{12}C$ and the
isoscalar quadrupole strength function in $^{12}C$ to compare with
inelastic $\alpha$ scattering data;
\item Explain the lifetime of $^{14}C$ (5730 years) which depends
sensitively on the nuclear wavefunction;
\item Calculate the properties of the Oxygen isotopes out to the
neutron drip line;
\item Solve for nuclei surrounding shell closure in order to
understand the origins shell closures.
\end{itemize}
Preliminary results for these goals are encouraging and will be
reported when we obtain additional results closer to convergence.

\section{Recent NCFC results}

Within the NCFC method, we adopt the harmonic oscillator (HO) single
particle basis which involves two parameters and we seek results
independent of those parameters either directly in a sufficiently
large basis or via extrapolation to the infinite basis limit.  The
first parameter $\hbar\Omega$ specifies the HO energy, the spacing
between major shells.  Each shell is labeled uniquely by the quanta of
its orbits $N = 2n + l$ (orbits are specified by quantum numbers
$n,l,j,m_j$) which begins with $0$ for the lowest shell and increments
in steps of unity.  Each unique arrangement of fermions (neutrons and
protons) within the HO orbits that is consistent with the Pauli
principle, constitutes a many-body basis state.  Many-body basis
states satisfying chosen symmetries are employed in evaluating the
Hamiltonian $H$ in that basis.  The second parameter is $N_{\max}$
which limits the total number of oscillator quanta allowed in the 
many-body basis states and thus limits the dimension D of the Hamiltonian
matrix.  $N_{\max}$ is defined to count the total quanta above the
minimum for the specific nucleus needed to satisfy the Pauli
principle.

\subsection{Quadrupole moment of $^6$Li}

Experimentally, $^6$Li has a surprisingly small quadrupole moment,
${\cal Q} = -0.083\,e\,{\rm fm}^2$.  This means that there are
significant cancellations between contributions to the quadrupole
moment from different parts of the wave function.  In
Ref.~\cite{Navratil01} the NCSM was shown to agree reasonably well
with the data using a nonlocal NN interaction.  Here we employ the
realistic JISP16 NN potential from inverse scattering that provides a
high-quality description of the NN data~\cite{Shirokov07}.  This NN
potential is sufficiently well-behaved that we may obtain NCFC results
for light nuclei~\cite{Maris09}.  Indeed, the left panel of Fig. 1
shows smooth uniform convergence of the ground state energy of $^6$Li.
The quadrupole moment does not exhibit the same smooth and uniform
convergence.
Nevertheless, the results for $15$ MeV$< \hbar\Omega < 25$ MeV where
the ground state energy is closest to convergence, strongly suggest a
quadrupole moment in agreement with experiment.
\begin{figure}[b]
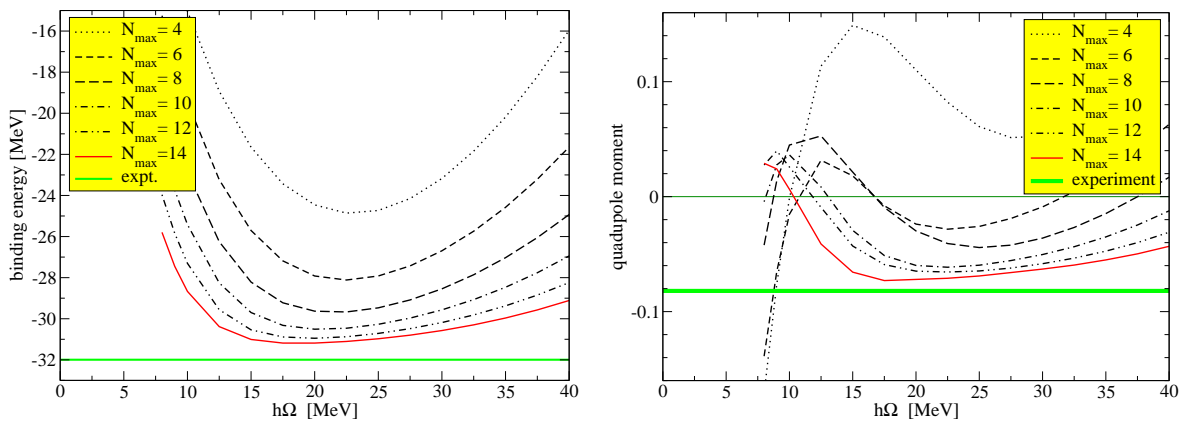

\includegraphics[width=18pc]{results_6Li.eps}\quad
\includegraphics[width=18pc]{results_6Li_Q.eps}
\caption{\label{label} Ground state energy of ${}^6$Li (left) and
quadrupole moment (right) obtained with the JISP16 NN potential as
function of $\hbar\Omega$ for several values of $N_{\max}$. }
\end{figure}

\subsection{Beryllium isotopes}

\begin{figure}[tb]
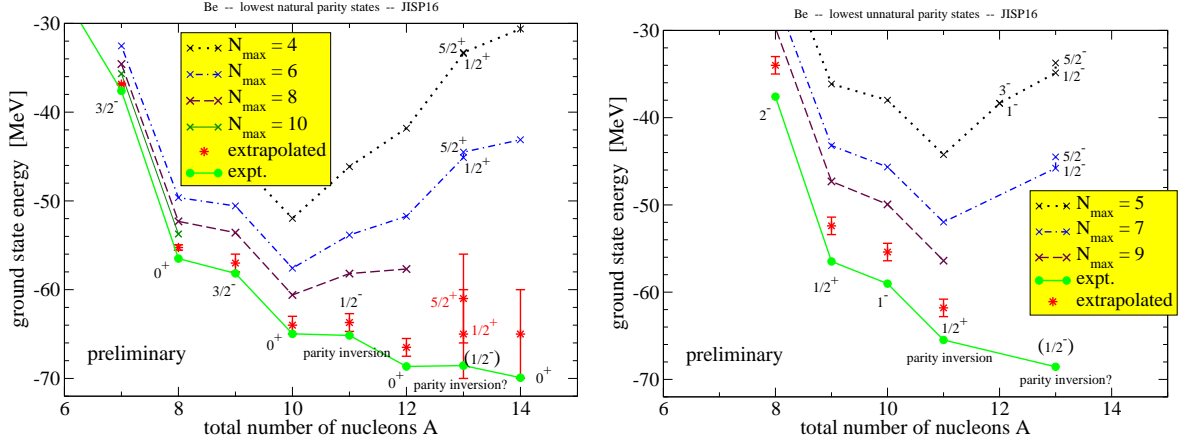

\includegraphics[width=18pc]{Be_natural.eps}\quad
\includegraphics[width=18pc]{Be_unnatural.eps}
\caption{\label{label}
Lowest natural (left) and unnatural (right) states of Be isotopes
obtained with the JISP16 NN potential.
Extrapolated results and their assessed
uncertainties are presented as red points
with error bars.
}
\end{figure}

Next, we use the Be isotopes to portray features of the physics
results and illustrate some of the computational challenges, again
with the JISP16 nonlocal NN potential \cite{Shirokov07}.  Fig. 2
displays the NCFC results for the lowest states of natural and
unnatural parity respectively in a range of Be isotopes along with the
experimental ground state energy.  Since our approach is guaranteed to
provide an upper bound to the exact ground state energy for the chosen
Hamiltonian, each point displayed for a particular value of $N_{\max}$
represents the minimum value as a function of $\hbar\Omega$ obtained
in that basis space.

In small model spaces, $N_{\max} = 4$ and $5$, the calculations
suggest a "U-shaped" curve for the binding energy as function of $A$,
the number of nucleons, whereas experimentally, the ground state
energy keeps decreasing with $A$, at least over the range in $A$ shown
in Fig. 2.  However, as one increases $N_{\max}$, the calculated
results get closer to the experimental values.  In order to obtain the
binding energies in the full configuration space (NCFC), we can use an
exponential fit to a set of results at 3 or 4 subsequent $N_{\max}$
values~\cite{Maris09}.  After performing such an extrapolation to
infinite model space, our calculated results follow the data very
closely as function of $A$.


In both panels of Fig. 2, the dimensions increase dramatically as one
increases $N_{\max}$ or increases atomic number $A$.  Detailed trends
will be presented in the next section.  The consequence is that
results terminate at lower $A$ as one increases $N_{\max}$.
Additional calculations are in process 
in order to reduce the uncertainties in the
extrapolations.

We note that the ground state spins are predicted correctly in all
cases except $A=11$ (and possibly at $A=13$, where the spin-parity of
the ground state is not well known experimentally) where there is a
reputed "parity inversion".  In this case a state of "unnatural
parity", one involving at least one particle moving up one HO shell
from its lowest location, unexpectedly becomes the ground state.
While Fig. 2 indicates we do not reproduce this parity inversion, we
do notice, however, that the lowest states of natural and unnatural
parity lie very close together.  This indicates that small corrections
arising from neglected effects, such as three-body (NNN) potentials
and/or additional contributions from larger basis spaces could play a
role here.  We are continuing to investigate these improvements.

\section{Computer Science and Applied Math challenges}

We outline the main factors underlying the need for improved
approaches to solving the nuclear quantum many-body problem as a large
sparse matrix eigenvalue problem.  Our goals require results as close
to convergence as possible in order to minimize our extrapolation
uncertainties. According to current experience, this requires the
evaluation of the Hamiltonian in as large a basis as possible,
preferably with $N_{\max} \ge 10$, in order to converge the ground
state wavefunction.

\begin{figure}[tb]
\begin{minipage}{20pc}
\includegraphics[width=20pc]{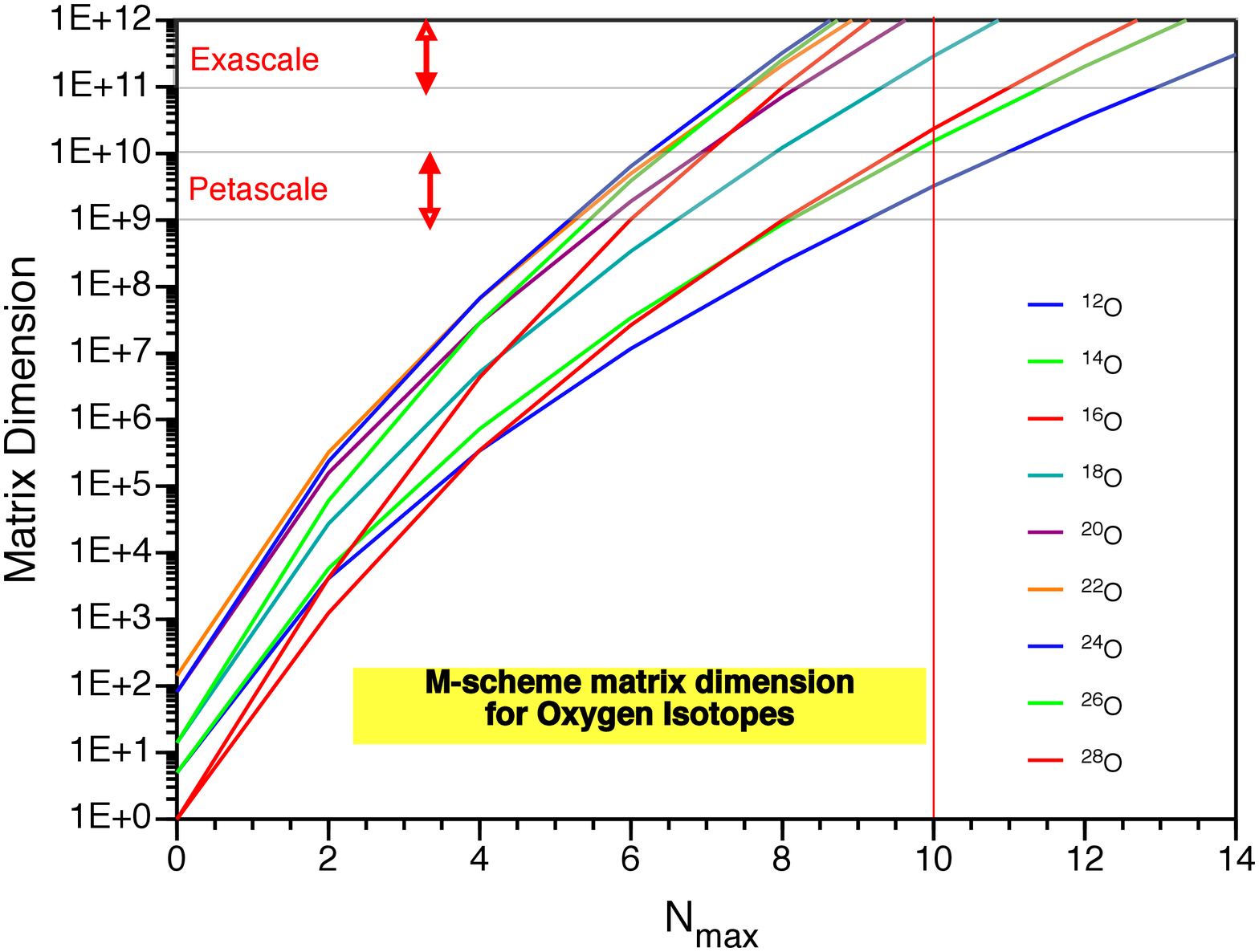}
\caption{\label{label}
Matrix dimension versus $N_{\max}$ \\
for stable and unstable Oxygen isotopes. \\
The vertical red line signals the boundary \\
where reasonable convergence emerges \\
on the right of it.}
\end{minipage}\hspace{0pc}%
\begin{minipage}{20pc}
\includegraphics[width=20pc]{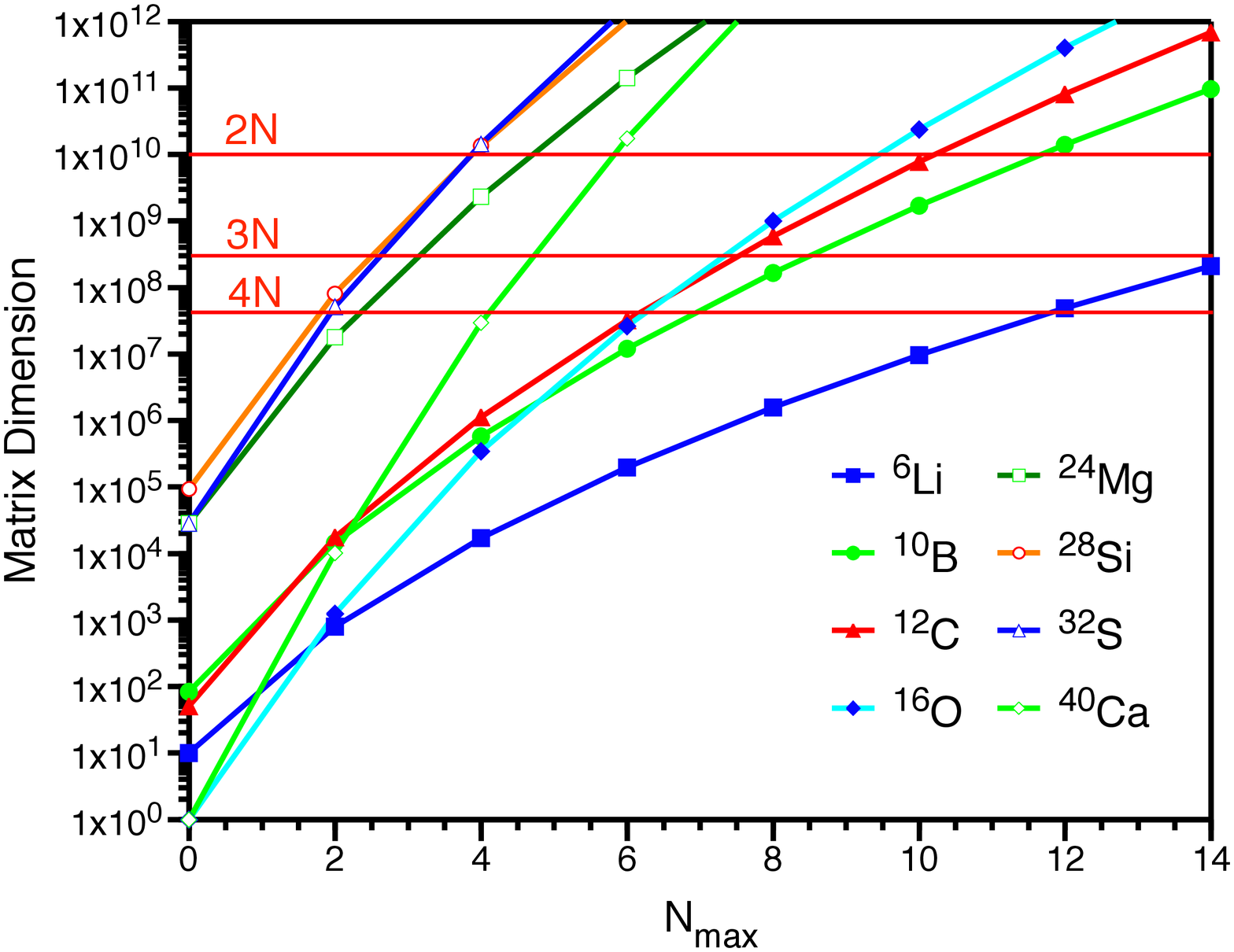}
\caption{\label{label}
Matrix dimension versus $N_{\max}$ \\
for a sample set of stable N=Z nuclei up to \\
$A=40$. Horizontal lines show expected limits\\
of Petascale machines for specific ranks of the \\
potentials (``2N"=NN, ``3N"=NNN, etc).
}
\end{minipage} 
\end{figure}
The dimensions of the Hamiltonian matrix grow combinatorially with
increasing $N_{\max}$ and with increasing atomic number $A$.  To gain a
feeling for that explosive growth, we plot in Fig. 3 the matrix
dimension (D) for a wide range of Oxygen isotopes.  In each case, we
select the "natural parity" basis space, the parity that coincides
with the lowest HO configuration for that nucleus.  The heavier of
these nuclei have been the subject of intense experimental
investigation and it is now believed that $^{28}O$ is not a
particle-stable nucleus even though it is expected to have a 
doubly-closed shell structure according to the phenomenological shell model.  
It would be very valuable to have converged {\it ab initio} NCFC results
for $^{28}O$ to probe whether realistic potentials are capable of
predicting its particle-unstable character.

We also include in Fig. 3 the estimated range that computer facilities
of a given scale can produce results with our current algorithms.  As
a result of these curves, we anticipate well converged NCFC results
for the first three isotopes of Oxygen will be achieved with Petascale
facilities since their curves fall near or below the upper limit of
Petascale at $N_{\max}=10$.

Dimensions of the natural parity basis spaces for another set of
nuclei ranging up to $A=40$ are shown in Fig. 4.  In addition, we
include estimates of the upper limits reachable with Petascale
facilities depending on the rank of the potential.  It is important to
note that theoretically derived 4N interactions are expected to be
available in the near future.  Though relatively less important than
2N and 3N potentials, their contributions are expected to grow
dramatically with increasing $A$.

\begin{figure}[tb]
\begin{minipage}{20pc}
\includegraphics[width=20pc]{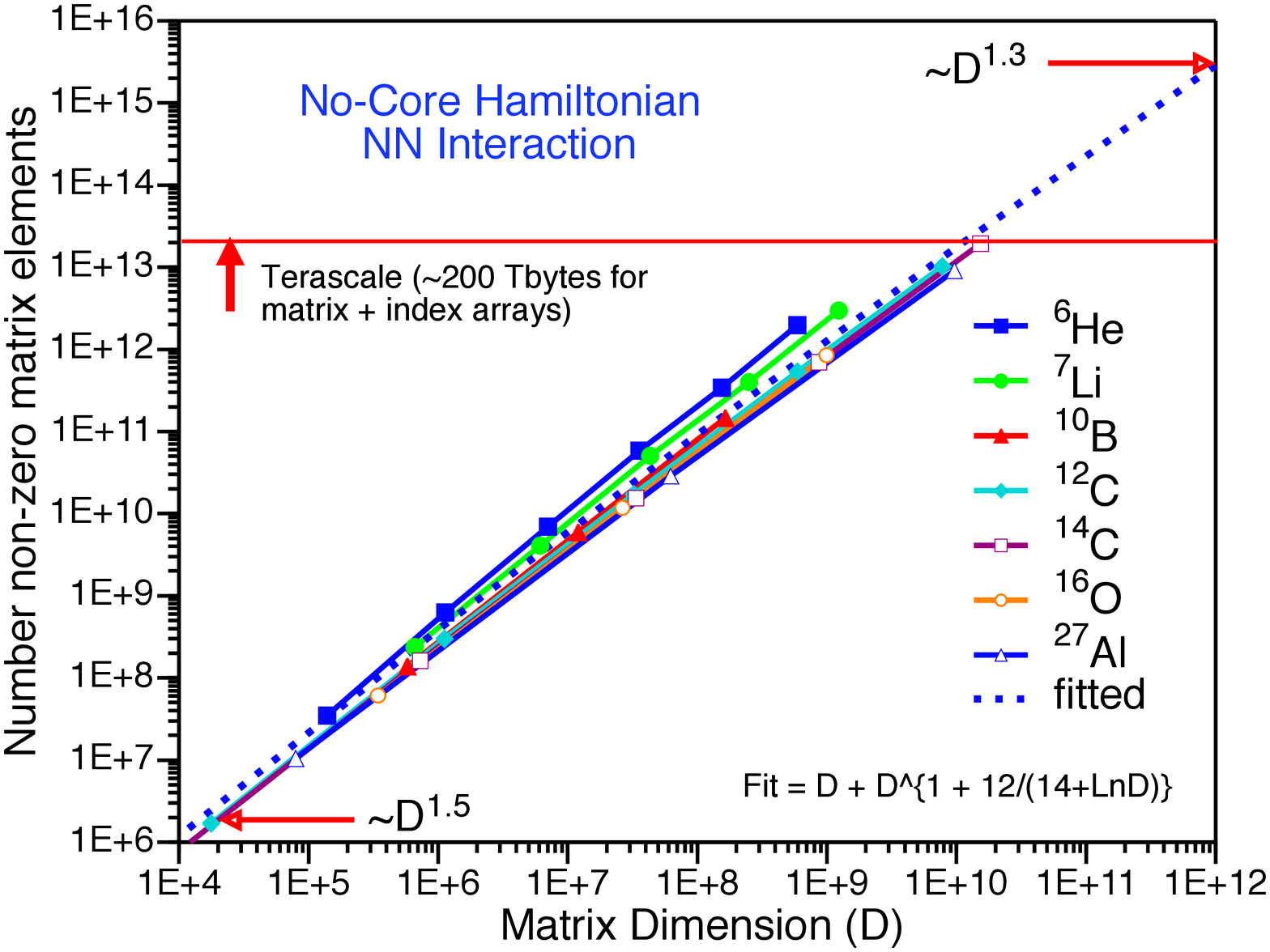}
\caption{\label{label}
Number of nonzero matrix \\
elements versus matrix dimension (D) \\
for an NN potential and for a selection \\
of light nuclei. The dotted line depicts a \\
function describing the calculated points.
}
\end{minipage}\hspace{0pc}%
\begin{minipage}{20pc}
\includegraphics[width=20pc]{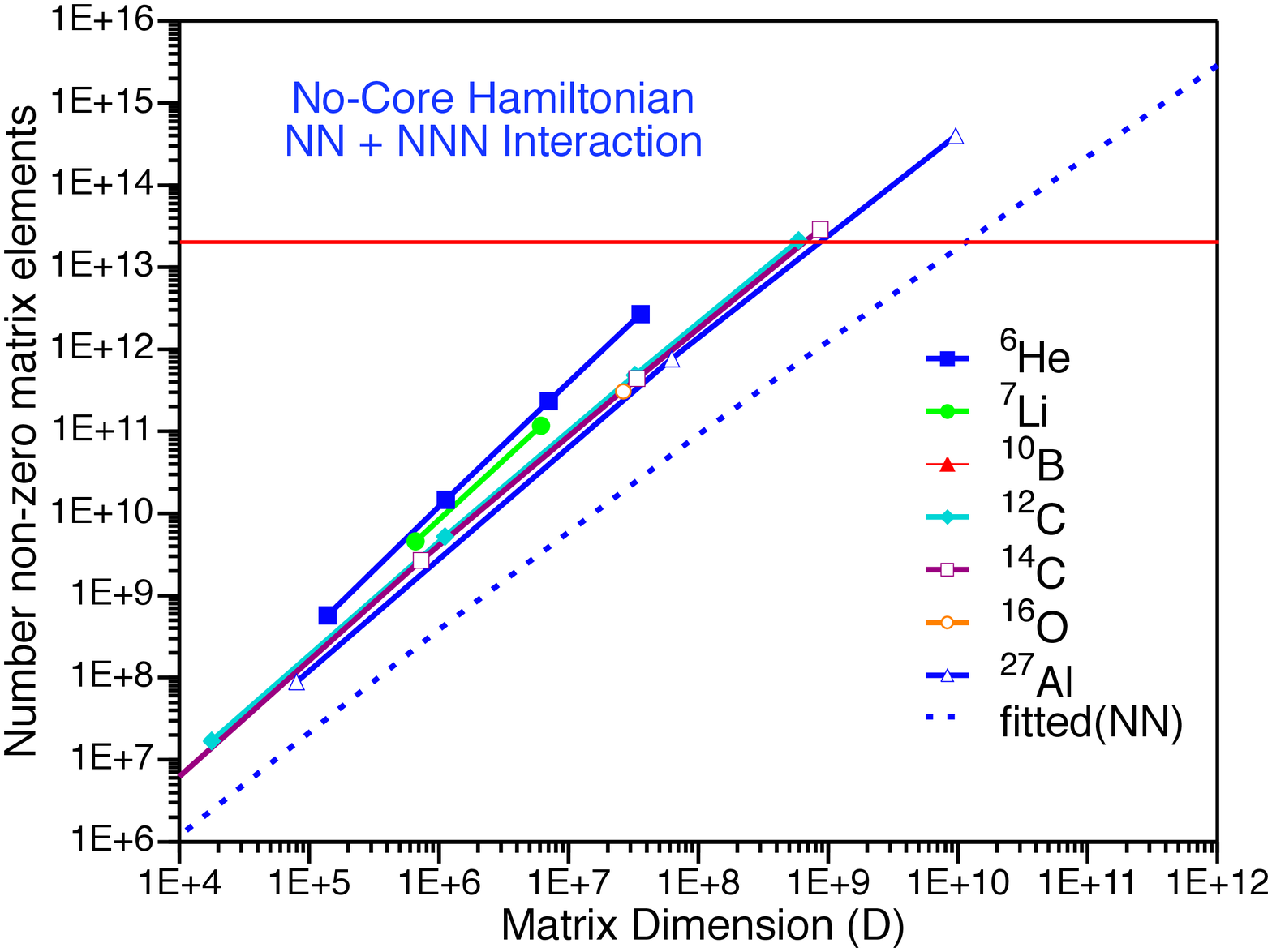}
\caption{\label{label}
Same as Fig. 5 but for an NNN \\
potential.  The NN dotted line of Fig. 5 \\
is repeated as a reference.  The NNN slope \\
and magnitude are larger than the NN case. \\
The NNN case is larger by a factor of 10-100.  
}
\end{minipage} 
\end{figure}

A significant measure of the computational burden is presented in
Figs. 5 and 6 where we display the number of non-zero many-body matrix
elements as a function of the matrix dimension (D).  These results are
for representative cases and show a useful scaling property.  For
Hamiltonians with NN potentials, we find a useful fit $F(D)$ for the
non-zero matrix elements with the function
\begin{eqnarray}
F(D) = D + D^{1 + \frac{12}{14 + {\ln D}}}.
\end{eqnarray}
The heavier systems displayed tend to be slightly below the fit while
the lighter systems are slightly above the fit.  The horizontal red
line indicates the expected limit of the Jaguar facility (150,000
cores) running one of these applications assuming all matrix elements
and indices are stored in core.  By way of contrast, we portray the
more memory-intensive situation with NNN potentials in Fig. 6, where we
retain the fitted curve of Fig. 5 for reference.  The horizontal red
line indicates the same limit shown in Fig. 5.

Looking forward to the advent of exascale facilities and the
development of 4N potentials, we present in Table 1 the matrix
dimensions and storage requirements for a set of light nuclei for
potentials ranging from 2N through 4N.  

\begin{table}
\begin{center}
\begin{tabular}{l|c|r|r|r|r}
nucleus    & $N_{\max}$ & dimension & 2-body & 3-body & 4-body  \\ 
\hline
${}^{6}$Li  & 12 & $4.9\cdot 10^6$    &   0.6 GB&          33 TB &   590 TB  \\
${}^{12}$C  &  8 & $6.0\cdot 10^8$    &    4 TB & 180 TB &   4 PB  \\
${}^{12}$C  & 10 & $7.8\cdot 10^9$    &   80 TB & 5 PB & 140 PB \\
${}^{16}$O  &  8 & $9.9\cdot 10^8$    &    5 TB & 300 TB &  5 PB \\
${}^{16}$O  & 10 & $2.4\cdot 10^{10}$ &  230 TB &  12 PB & 350 PB \\
%
%
%
%
\hline
%
${}^8$He    & 12 & $4.3\cdot 10^{8}$   &     7 TB &  300 TB &  7 PB \\
%
${}^{11}$Li & 10 & $9.3\cdot 10^8$     &    11 TB & 390 TB &  10 PB \\ 
%
${}^{14}$Be &  8 & $2.8\cdot 10^9$     &    32 TB & 1100 TB & 28 PB \\ 
%
%
${}^{20}$C  &  8         &$2\cdot 10^{11}$&   2 PB & 150 PB & 6 EB   \\ 
%
${}^{28}$O  &  8 & $1\cdot 10^{11}$    &    1 PB &  56 PB  & 2 EB  \\ 
\end{tabular}
\end{center}
\caption{Table of storage requirements of current 
version of MFDn for a range of applications.  
Roughly speaking, 
entries up to 400TB imply Petascale while 
entries above 1PB imply Exascale
facilities will likely be required.}
\end{table}

\section{Algorithms of MFDn}

Our present approach embodied in the code Many Fermion Dynamics -
nuclear (MFDn) \cite{Vary1,Vary2,Sternberg08} involves several key
steps:
\begin{enumerate}
\item enumerate the single-particle basis states for the neutrons and
the protons separately with good total angular momentum projection
$m_j$,
\item enumerate and store the many-body basis states subject to
user-defined constraints and symmetries such as $N_{\max}$, parity and
total angular momentum projection $M$,
\item evaluate and store the many-nucleon Hamiltonian matrix in this
many-body basis using input files for the NN (and NNN interaction if
elected),
\item obtain the lowest converged eigenvalues and eigenvectors using
the Lanczos algorithm with a distributed re-orthonormalization scheme,
\item transform the converged eigenvectors from the Lanczos vector
space to the original basis and store them in a file,
\item evaluate a suite of experimental observables using the stored
eigenvectors.
\end{enumerate}
All these steps, except the first, which does not require any
significant amount of CPU time, are parallelized using MPI.  Fig. 7
presents the allocation of processors for computing and storing the
many-body Hamiltonian on $n(n+1)/2$ processors.  (We illustrate the
allocation with 15 processors so the pattern is clear for an arbitrary
value of $n$.)  We assume a symmetric Hamiltonian matrix and
compute/store only the lower triangle.

In step two, the many-body basis states are round-robin distributed
over the $n$ diagonal processors.  Each vector is also distributed
over $n$ processors, so we can (in principle) deal with arbitrary
large vectors.  Because of the round-robin distribution of the basis
states, we obtain excellent load-balancing; however, the price we pay
is that any structure in the sparsity pattern of the matrix is lost.
We have implemented multi-level blocking procedures to enable an
efficient determination of the non-zero matrix elements to be
evaluated.  These procedures have been presented recently in
Ref. \cite{Sternberg08}.

After the determination and evaluation of the non-zero matrix
elements, we solve for the lowest eigenvalues and eigenvectors using
an iterative Lanczos algorithm, step four.  The matvec operation, and
its communication patterns, for each Lanczos iteration is shown in
Fig. 8 where two sets of operations account for the storage of only
the lower triangle of the matrix.

\begin{figure}[tb]
\begin{minipage}{20pc}
\includegraphics[width=20pc]{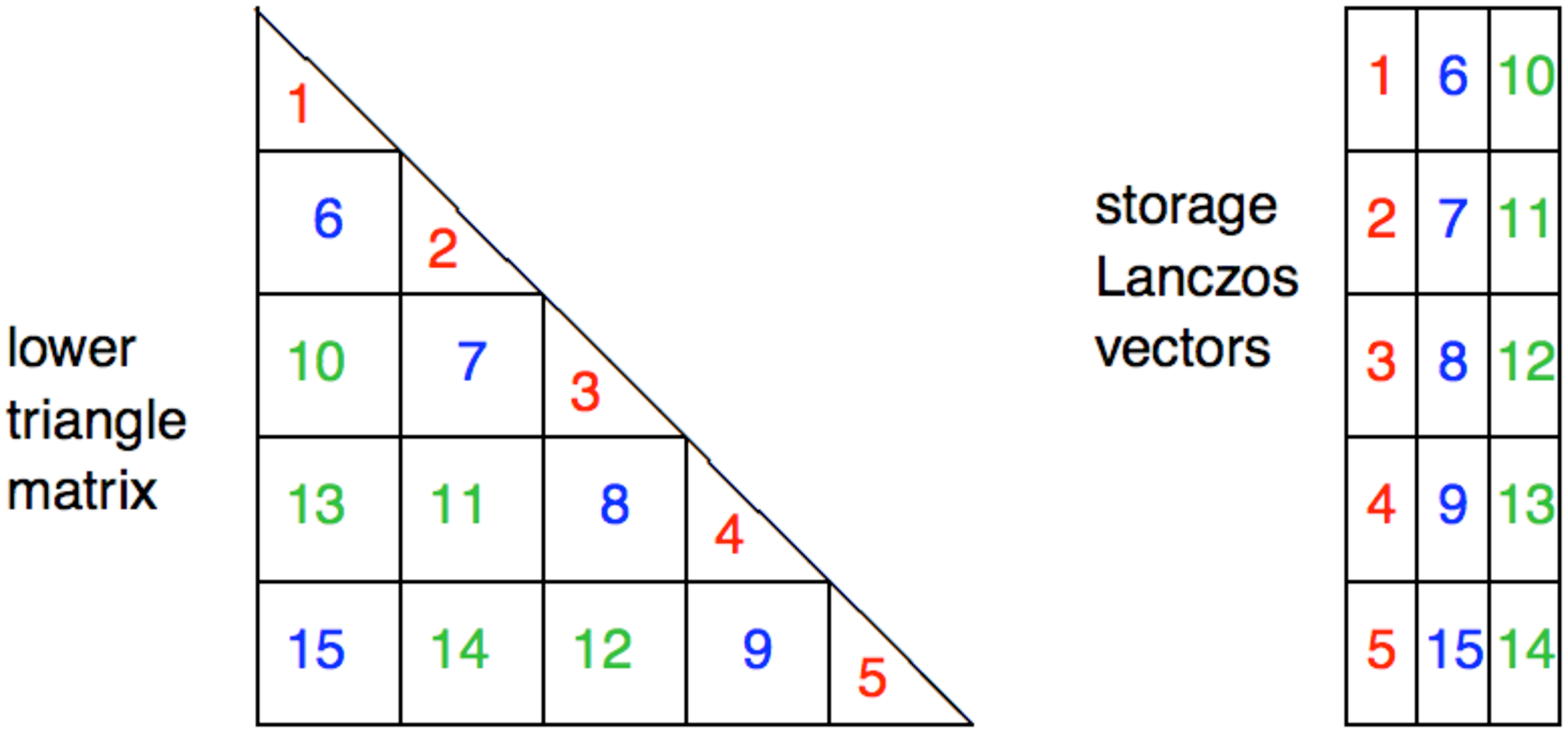}
\caption{\label{label}
MFDn scheme for distributing the \\
symmetric Hamiltonian matrix over $n(n+1)/2$ \\
PE's and partitioning the Lanczos vectors. We \\
use $n=5$ "diagnonal" PE's in this illustration.
}
\end{minipage}\hspace{0pc}%
\begin{minipage}{20pc}
\includegraphics[width=20pc]{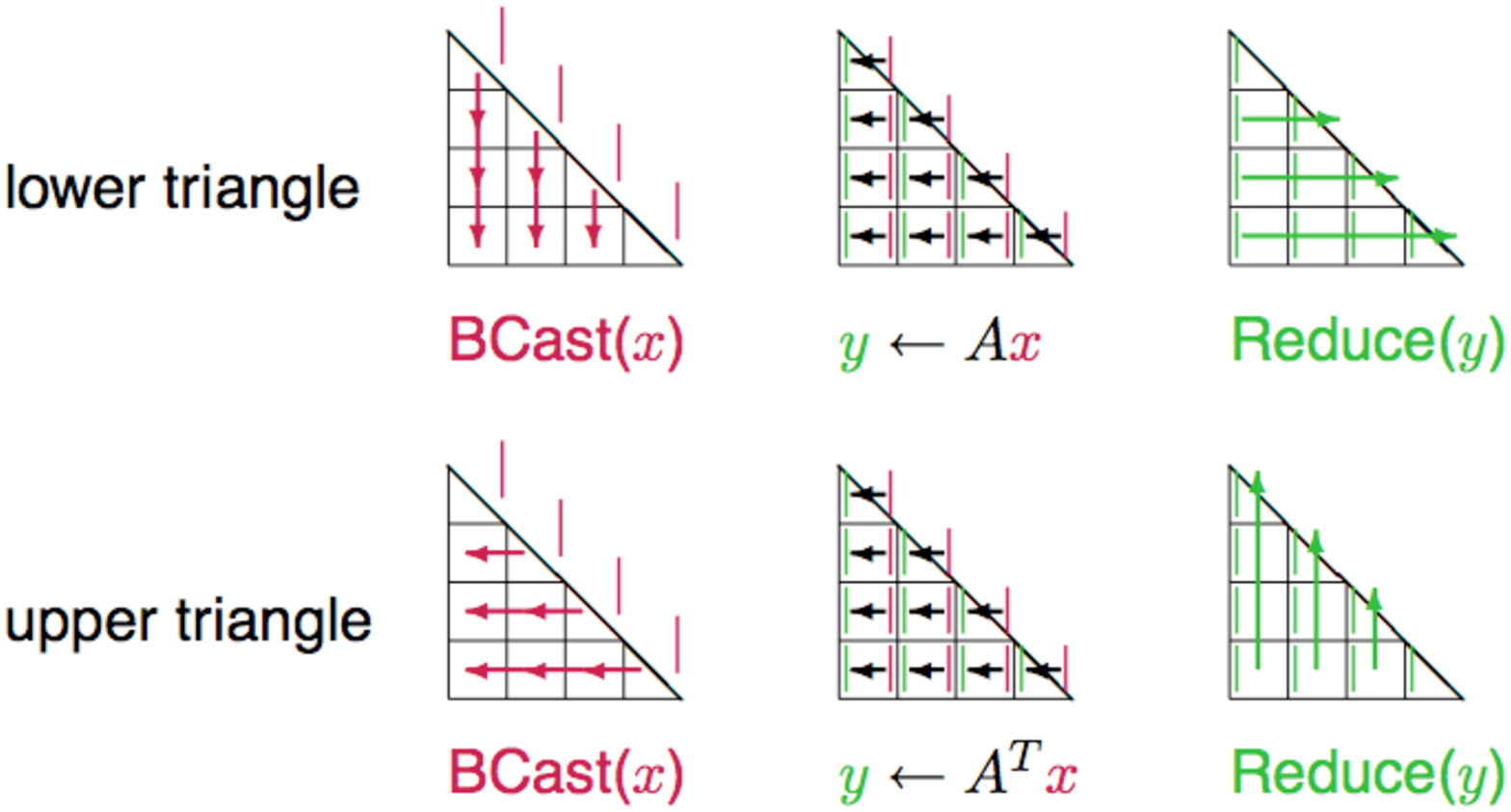}
\caption{\label{label}
MFDn scheme for performing the \\
Lanczos iterations based on the Hamiltonian \\
matrix stored in $n(n+1)/2$ PE's.  We use \\
$n=4$ "diagnonal" PE's in this illustration.
}
\end{minipage} 
\end{figure}

MFDn also employs a distributed re-orthonormalization technique
developed specifically for the parallel distribution shown in Fig. 7.
After the completion of the matvec, the new elements of the
tridiagonal Lanczos matrix are calculated on the diagonals.  The
(distributed) normalized Lanczos vector is then broadcast to all
processors for further processing.  Each previously computed Lanczos
vector is stored on one of $(n+1)/2$ groups of $n$ processors.  Each
group receives the new Lanczos vector, computes the overlaps with all
previously stored vectors within that group, and constructs a
subtraction vector, that is, the vector that needs to be subtracted
from the Lanczos vector in order to orthogonalize it with respect to the
Lanczos vectors stored in that group.  These subtraction vectors are
accumulated on the $n$ diagonal processors, where they are subtracted
from the current Lanczos vector.  Finally, the orthogonalized Lanczos
vector is re-normalized and broadcast to all processors for
initiating the next matvec.  One group is designated to also store
that Lanczos vector for future re-orthonormalization operations.

The results so far with this distributed re-orthonormalization appear
stable with respect to the number of processors over which the Lanczos
vectors are stored.  We can keep a considerable number of Lanczos vectors
in core, since all processors are involved in the storage, and each
part of a Lanczos vector is stored on one and only one processor.
For example, we have run 5600 Lanczos iterations on a matrix with
dimension 595 million and performed the distributed
re-orthonormalization using 80 groups of stored Lanczos vectors.  We
converged about 450 eigenvalues and eigenvectors.  Tests of the
converged states, such as measuring a set of their symmetries, showed
they were reliable.  No converged duplicate eigenstates were
generated.  Test cases with dimensions of ten to forty million
have been run on various numbers of processors to verify the strategy
produces stable results independent of the number of processors.  A
key element for the stability of this procedure is that the overlaps
and normalization are calculated in double precision, even though the
vectors (and the matrix) are stored in single precision.  Furthermore,
there is of course a dependence on the choice of initial pivot vector;
we favor a random initial pivot vector.

\section{Accelerating convergence}

We are constantly seeking new ideas for accelerating convergence as
well as saving memory and CPU time.  The situation is complicated by
the wide variety of strong interaction Hamiltonians that we employ and
we must simultaneously investigate theoretical and numerical methods
for "renormalizing" (softening) the potentials.  Softening an
interaction reduces the magnitude of many-body matrix elements between
very different basis states such as those with very different total
numbers of HO quanta. However, all methods to date soften the
interactions at the price of increasing their complexity.  For
example, starting with a strong NN interaction, one renormalizes it to
improve the convergence of the many-body problem with the softened NN
potential \cite{Bogner:2007rx} but, in the process, one induces a new
NNN potential that is required to keep the final many-body results
invariant under renormalization.  In fact, 4N potentials and higher
are also induced\cite{Navratil07}.  Given the computational cost of 
these higher-body
potentials in the many-body problem as described above, it is not
clear how much one gains from renormalization - for some NN
interactions it may be more efficient to attempt larger basis spaces
and avoid the renormalization step.  This is a current area of intense
theoretical research.

In light of these features of strongly interacting nucleons and the
goal to proceed to heavier nuclei, it is important to investigate
methods for improving convergence.  A number of promising methods have
been proposed and each merits a significant research effort to assess
their value over a range of nuclear applications.  A sample list
includes:
\begin{itemize}
\item Symplectic no core shell model (SpNCSM) - extend basis spaces
beyond their current $N_{\max}$ limits by adding relatively few
physically-motivated basis states of symplectic symmetry which is a
favored symmetry in light nuclei
\cite{Dytrych}
\item Realistic single-particle basis spaces \cite{Negoita08}
\item Importance truncated basis spaces \cite{Roth07}
\item Monte Carlo sampling of basis spaces \cite{Ormand94,Honma96}
\item Coupled total angular momentum $J$ basis space
\end{itemize}
Several of these methods can easily be explored with MFDn.  For example, 
while MFDn
has some components that are specific for the HO basis, it is
straightforward to switch to another single-particle basis with the
same symmetries but with different radial wavefunctions.  Alternative
truncation schemes such as Full Configuration Interaction (FCI) basis
spaces are also implemented and easily switched on by appropriate
input data values. These flexible features have proven convenient for
addressing a wide range of physics problems.

There are additional suggestions from colleagues that are worth
exploring as well such as the use of wavelets for basis states.  These
suggestions will clearly involve significant research efforts to
implement and assess.

\section{Performance of MFDn}

We present measures of MFDn performance in Figs. 9 and 10.  Fig. 9
displays the performance characteristics of MFDn from the beginning of
the SciDAC-UNEDF program to the present, roughly a 2 year time span.
Many of the increments along the path of improvements have been
documented \cite{Sternberg08,Sosonkina08,Laghave09}.

\begin{figure}[tb]
\begin{minipage}{20pc}
\includegraphics[width=20pc]{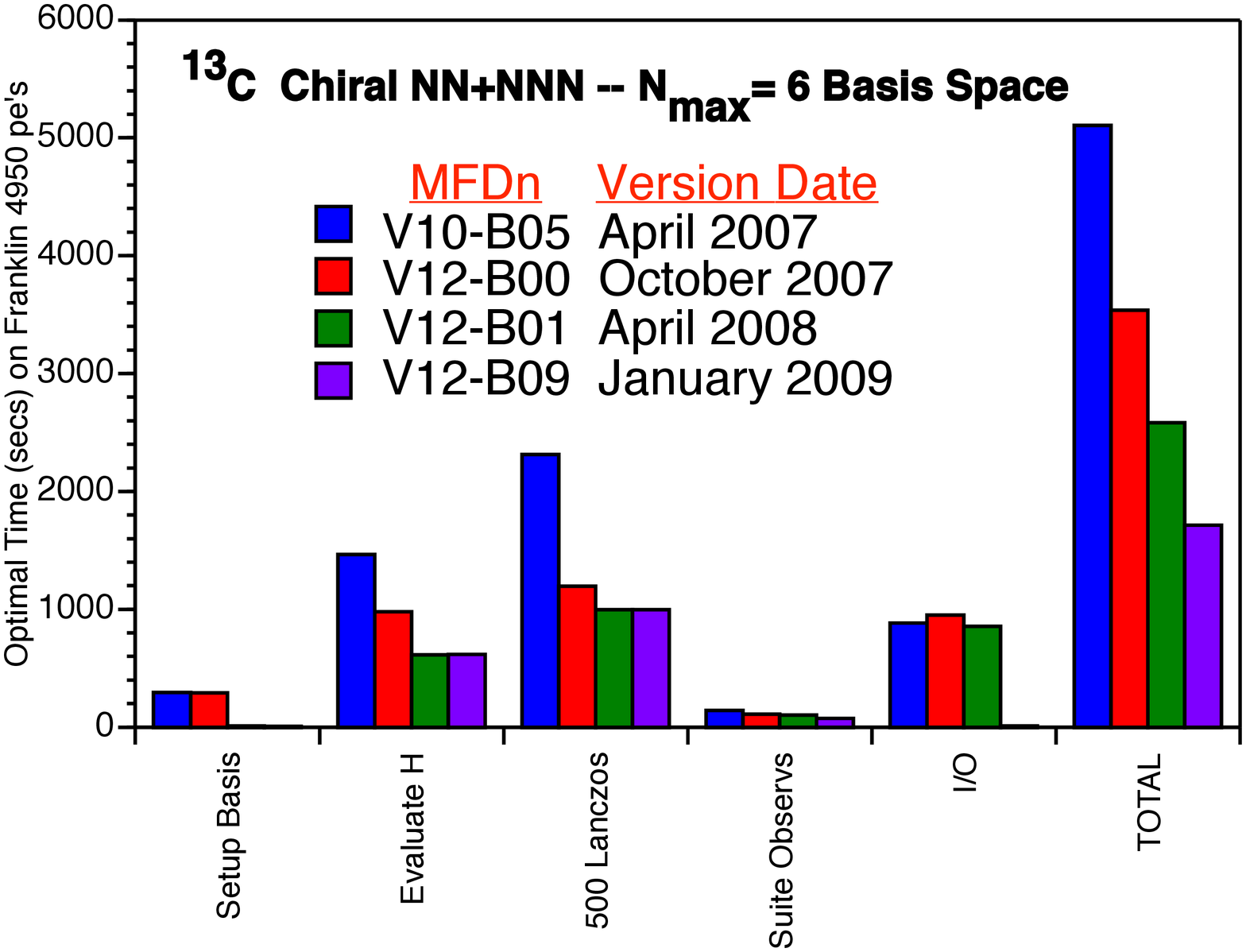}
\caption{\label{label}
Performance improvement of \\
MFDn over a 2 year span displayed for \\
major sections of the code.
}
\end{minipage}\hspace{0pc}%
\begin{minipage}{20pc}
\includegraphics[width=20pc]{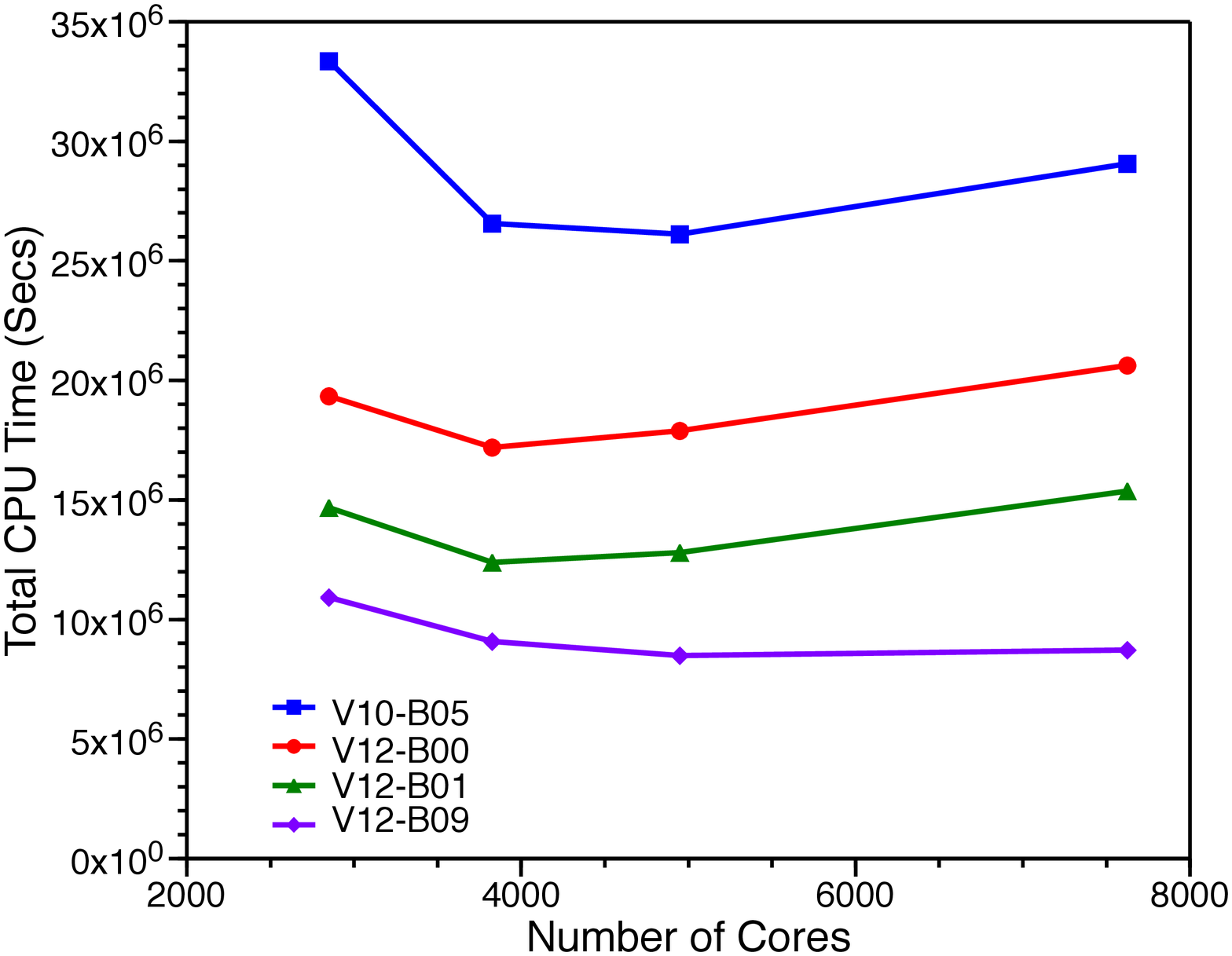}
\caption{\label{label}
Total CPU time for the same test \\
cases evaluated in Fig. 9 running on various \\
numbers of Franklin cores.
}
\end{minipage} 
\end{figure}

Fig. 9 displays the timings for major sections of the code as well as
the total time for a test case in $^{13}$C using a NNN potential on
4950 Franklin processors.  The timings are specified for MFDn versions that
have emerged during the SciDAC-UNEDF program.  The initial version
tested, V10-B05, represents a reasonable approximation to the version
prior to SciDAC-UNEDF.  We note that the overall speed has improved by
about a factor of 3.  Additional effort is needed to further reduce
the time to evaluate the many-body Hamiltonian H and to perform the
matvec operations as the other sections of the code have undergone
dramatic time reductions.

We present in Fig. 10 one view of the strong scaling performance of
MFDn on Franklin.  Here the test case is held fixed at the case shown
in Fig. 9 and the number of processors is varied from 2850 to 7626.
Due to memory limitations associated with the very large input NNN
data file (3~GB), scaling is better than ideal at (relatively) low
processor numbers.  At the low end, there is significant redundancy of
calculations since the entire NNN file cannot fit in core and must be
processed in sections.  As one increases the number of processors,
more memory becomes available for the NNN data and less redundancy is
incurred.  At the high end, the scaling begins to show signs of
deteriorating due to increased communications time relative to
computations.  The net result is the dip in the middle which we can
think of as the "sweet spot" or the ideal number of processors for
this application.  Clearly, this implies significant preparation work
for large production runs to insure maximal efficiency in the use of
limited computational resources.  For the largest feasible
applications, this also implies there is high value to finding a
solution for the redundant calculations brought about by the large
input files.  We are currently working on a hybrid OpenMP/MPI version
of MFDn, which would significantly reduce this redundancy.  However,
at the next model space, the size of the input file increases to
33~GB, which we need to process in sections on Franklin.

\section{Conclusions and Outlook}

Microscopic nuclear structure/reaction physics is enjoying a
resurgence due, in large part, to the development and application of
{\it ab initio} quantum many-body methods with realistic strong
interactions.  True predictive power is emerging for solving
long-standing fundamental problems and for influencing future
experimental programs.

We have focused on the no core shell model (NCSM) and the no core full
configuration (NCFC) approaches and outlined our recent progress.
Large scale applications are currently underway using leadership-class
facilities and we expect important progress on the path to detailed
understanding of complex nuclear phenomena.

We have also outlined the challenges that stand in the way of further
progress.  More efficient use of memory, faster I/O and faster
eigensolvers will greatly aid our progress.

\ack This work was supported in part by DOE grants DE-FC02-09ER41582
and DE-FG02-87ER40371.  Computational resources are provided by DOE
through the National Energy Research Supercomputer Center (NERSC) and
through an INCITE award (David Dean, PI) at Oak Ridge National Laboratory and 
Argonne National Laboratory.

\end{document}